\documentclass[reprint,
onecolumn, 
superscriptaddress,
showkeys,
 amsmath,amssymb,
 aps,
 prf,
]{revtex4-2}
\usepackage{graphicx}
\usepackage{amsmath}
\usepackage{dcolumn}
\usepackage{bm}
\usepackage[dvipsnames]{xcolor}
\usepackage{caption}
\usepackage{subcaption}
\begin{document}

\title{Taylor's swimming sheet near a soft boundary}
\author{Aditya Jha}
	\email{aditya.jha@u-bordeaux.fr}
 \affiliation{Univ. Bordeaux, CNRS, LOMA, UMR 5798, F-33405 Talence, France.}
\author{Yacine Amarouchene}
\affiliation{Univ. Bordeaux, CNRS, LOMA, UMR 5798, F-33405 Talence, France.}
\author{Thomas Salez}
	\email{thomas.salez@cnrs.fr}
\affiliation{Univ. Bordeaux, CNRS, LOMA, UMR 5798, F-33405 Talence, France.}

\begin{abstract}
In 1951, G.I. Taylor modeled swimming microorganisms by hypothesizing an infinite sheet in 2D moving in a viscous medium due to a wave passing through it. This simple model not only captured the ability of microorganisms to swim due to the wavy motion of a flagella, but further development into the model captured the optimal nature of metachronal waves observed in ciliates. While the additional effects of nearby rigid boundaries and complex environments have been addressed, herein we explore the correction induced by a nearby soft boundary. Our simple model allows us to show that the magnitude of the swimming velocity gets modified near soft boundaries, and reduces for transverse waves while it increases for longitudinal waves. We further delve into the energetics of the process and the deformation of the corresponding soft boundary, highlighting the synchronization of the oscillations induced on the soft boundary with the waves passing through the sheet and the corresponding changes to the power exerted on the fluid. 

\end{abstract}
\maketitle

\section{\label{sec:intro}Introduction}

The motility of microorganisms controls their access to food and nutrients and hence, is primitive to their survival~\cite{lauga2009hydrodynamics,lauga2020fluid}. In spermatozoan cells, this basic task is fundamental to reproduction~\cite{smelser1974swimming, shukla1978swimming, suarez2010sperm}. Given the vast variability in the biological characteristics of different microorganisms, different strategies are employed to achieve mobility~\cite{childress1981mechanics}. However, these strategies are in stark contrast to the ones of larger animals. The difference arises out of the relative importance of inertia across various scales. While larger animals can move by imparting momentum to the media around them, viscous dissipation at the small microscopic scales deprives the microorganisms of this strategy~\cite{happel2012low,stone2015low}. 

As viscous dissipation becomes increasingly important, time-reversal symmetry forbids reciprocal mobility strategies. This `scallop theorem' represents a fundamental characteristic of mobility in viscous media, first formalized by Purcell in 1977~\cite{purcell2014life, lauga2011life, ishimoto2012coordinate}. However, an easy solution commonly observed for many flagellated and ciliated microorganisms is the presence of propagating waves on their appendages that break symmetry and lead to propulsion~\cite{brennen1977fluid}. The variability in the properties of the wave, including its amplitude, wavelength, and frequency, generates a large variety of propulsion speeds. 

In an attempt to understand the details of propulsion of these flagellated microorganisms, Taylor~\cite{taylor1951analysis} introduced a simplified model aimed to capture the essence of their mobility. This model, which consisted of an infinite sheet in a highly viscous medium with a passing wave, captured the essential relationship between the characteristics of the wave and the propulsion velocity achieved. Further research around this model has looked into the added effects of confinement by rigid walls~\cite{katz1974propulsion,reynolds1965swimming,fauci1995sperm}, inertia of the fluid~\cite{tuck1968note}, complex media such as liquid crystals~\cite{krieger2014locomotion,krieger2015microscale,soni2023taylor}, viscoelastic fluids and activity~\cite{velez2013waving,riley2014enhanced}. Transient effects were also explored~\cite{pak2010transient}, and numerical calculations provided details on the large-amplitude and complex waveforms required for optimal transport~\cite{montenegro2014optimal}. Subsequently, the energetics of synchronization of multiple sheets was explored~\cite{chrispell2013actuated,liao2021energetics} in order to understand the phase locking of multiple flagellas. 

Besides, lubrication flows near soft surfaces have shown how novel forces emerge when the elasticity of the substrate is taken into account~\cite{sekimoto1993mechanism, skotheim2004soft, bureau2023lift,rallabandi2024fluid}. These forces arising out of soft lubrication remain coupled to the nature of the elastic solid bounding the flow and have been experimentally confirmed with different substrates~\cite{zhang2020direct,bertin2021contactless}. Previous research~\cite{trouilloud2008soft} has shown that the inclusion of wall elasticity leads to locomotion even in the presence of reciprocal swimming strategies, highlighting the importance of understanding the effects of wall elasticity for swimming -- especially as soft boundaries are ubiquitous in microbiology. While theoretical research based on point-like swimmers has shown increased speeds when confined in elastic tubes~\cite{ledesma2013enhanced}, the swimming-sheet model has shown decreased speeds near passive membranes~\cite{dias2013swimming}. In either case, a detailed understanding of the energetics of the problem and the deflection of the elastic boundary remains to be done. On the experimental front, a study of swimming bacteria~\cite{tchoufag2019mechanisms} has demonstrated a decrease in the swimming velocity with increasing softness. 

The current work aims to explore the details of swimming near a soft elastic boundary by focusing on Taylor's swimming sheet near a Winkler elastic solid~\cite{dillard2018review}. This simplified approach lets us elaborate on the modifications induced to the swimming velocity due to a single elastic parameter and provides explanations and criteria for increasing or decreasing the swimming velocity. Moreover, it allows for a precise study of the changes in energy consumption and the deformation of the elastic boundary itself. The remainder of this article is organized as follows. We start by describing the swimming-sheet problem, followed by the perturbative decomposition for small-amplitude waves passing through the sheet. The perturbative approach drastically simplifies the problem, allowing us to calculate the velocity of the swimming sheet. Furthermore, we address the energetics of the problem and the deformation induced in the substrate, helping us to characterize the synchronization of the swimming sheet with the elastic boundary. 

\section{\label{sec:model}Model}
\subsection{Governing equations}
\begin{figure}[h]
\begin{center}
\includegraphics[width=8cm]{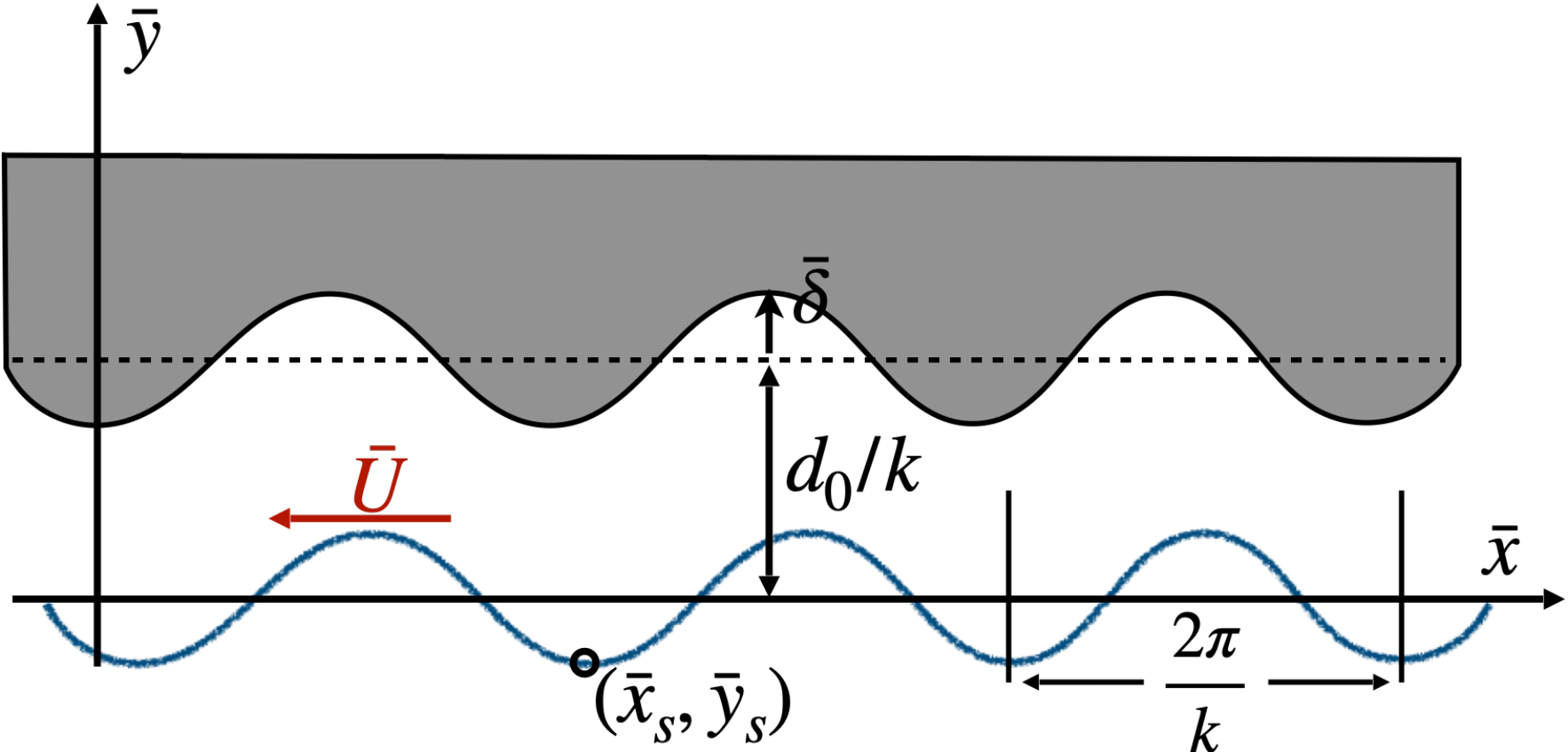}
\caption{Schematic of the 2D system in the lab frame of reference. An infinite sheet (blue line) is located a mean distance of $d_0/k$ from a no-slip deformable boundary (grey). The prescribed time-dependent deformation of the sheet is given by Eqs.~(\ref{eq:material_point_xs}-\ref{eq:material_point_ys}), which correspond to a wave passing through it with angular wavenumber $k$ and angular frequency $\omega$. The traveling wave causes the sheet to move with a velocity $\bar{U}$ in the opposite direction, as depicted.   }
\label{fig:scheme}
\end{center}
\end{figure}

We assume an infinite sheet in 2D with a wave passing through it, where the oscillations along time $\bar{t}$ occur at angular frequency $\omega$ with angular wavenumber $k$, as shown schematically in Fig.~\ref{fig:scheme}. The passing wave leads to the sheet's motion with a velocity $\bar{U}$ towards the left. To simplify the calculations, we shift to the swimmer's frame of reference. The position of a material point of the sheet $(\bar{x}_{\textrm{s}},\bar{y}_{\textrm{s}})$ is assumed to be described by the periodic waving motion as: 
\begin{align}
        \bar{x}_{\textrm{s}}(\bar{x},\bar{t}) &= \bar{x} + A \cos(k\bar{x}-\omega \bar{t}-\bar{\phi}), \label{eq:material_point_xs} \\
    \bar{y}_{\textrm{s}}(\bar{x},\bar{t}) &= B \sin(k\bar{x}-\omega \bar{t}), \label{eq:material_point_ys} 
\end{align}
where $A$ and $B$ denote the amplitudes of the longitudinal and transverse waving motions and $\bar{\phi}$ denotes the phase difference between these two waving motions. 

Assuming the irrelevance of fluid inertia to describe the microswimmer's motion, Stokes' equation governs the motion of the fluid. Alongside the incompressibility condition, the dynamics of the fluid is then given by the following equations: 
\begin{align}
    \nabla.\bm{\bar{\sigma}} &= \eta\nabla^2\Vec{\bar{u}}-\nabla \bar{p} = 0,\\
    \nabla.\Vec{\bar{u}} &= 0, 
\end{align}
where $\bm{\bar{\sigma}}(\bar{x},\bar{y},\bar{t})$ denotes the stress tensor, $\bar{p}(\bar{x},\bar{y},\bar{t})$ denotes the excess fluid pressure with respect to the static atmospheric one, $ \Vec{\bar{u}}(\bar{x},\bar{y},\bar{t})=[\bar{u}_{\bar{x}}(\bar{x},\bar{y},\bar{t}),\bar{u}_{\bar{y}}(\bar{x},\bar{y},\bar{t})]$ denotes the fluid velocity, and $\eta$ denotes the fluid viscosity. The vertical deflection $\bar{\delta}(\bar{x},\bar{t})$ of the soft boundary with respect to its flat rest state is assumed to be described by Winkler's model: 
\begin{align}
    \bar{\delta}(\bar{x},\bar{t}) = \frac{L}{2G}\left[\bar{p}(\bar{x},d_0/k,\bar{t})-2\frac{\partial \bar{u}_{\bar{y}}}{ \partial \bar{y} }(\bar{x},d_0/k,\bar{t})\right], \label{eq:winkler}
\end{align}
where $G$ denotes the shear modulus of the soft solid, and $L$ is its thickness. The horizontal deformation of the soft solid is null, owing to the specific choice for the elastic response. 

For the viscous fluid in between the waving sheet and the soft boundary, a no-slip boundary condition is imposed at both the sheet and the deformable boundary. Thus, the velocity of the fluid at the waving sheet is described by the velocity of the material points, as: 
\begin{align}
    \Vec{\bar{u}}(\bar{x}_{\textrm{s}},\bar{y}_{\textrm{s}},\bar{t}) = \frac{\textrm{d}}{\textrm{d}\bar{t}}(\bar{x}_{\textrm{s}},\bar{y}_{\textrm{s}}),
\end{align}
and the velocity of the fluid at the soft wall is given by:
\begin{align}
    \Vec{\bar{u}}(\bar{x},d_0/k+\bar{\delta},\bar{t}) = \left[\bar{U}, \frac{\textrm{d}\bar{\delta}}{\textrm{d}\bar{t}}(\bar{x},\bar{t})\right],
\end{align}

\subsection{Non-dimensionalization}
We non-dimensionalize the equations by using $\omega^{-1}$ and $k^{-1}$ as scales for time and length. The characteristic scale of the pressure comes is $\eta \omega$. We introduce the parameter $\epsilon = (A^2k^2+B^2k^2)^{1/2}$. The non-dimensional longitudinal and transverse waving amplitudes thus become $a = Ak/\epsilon$ and $b = Bk/\epsilon$. The non-dimensional quantities are denoted with the overbars removed, leading to the following dimensionless equations for the motion of the sheet: 
\begin{align}
    x_{\textrm{s}}(x,t) &= x + \epsilon a \cos(x-t-\phi), \\
    y_{\textrm{s}}(x,t) &= \epsilon b \sin(x-t), 
\end{align}
where $\phi=\bar{\phi}$. The non-dimensional deflection of the sheet is then denoted as:
\begin{align}
    \delta(x,t) = \kappa \left[p(x,d_0,t)-2\frac{\partial u_y}{\partial y}(x,d_0,t)\right],  \label{eq:winkler_nondim}
\end{align}
where $\kappa =Lk\eta \omega/(2G)$ is the dimensionless elastic compliance. Since the problem is two-dimensional, we introduce the stream function $\psi(x,y,t)$, defined by: 
\begin{align}
    u_x &= \frac{\partial \psi}{\partial y}, \\
    u_y &= -\frac{\partial \psi}{\partial x}. 
\end{align}
The Stokes' equation then implies that the stream function is the solution of the biharmonic equation: 
\begin{align}
    \nabla^4\psi = 0.
\end{align}
Besides, the pressure is related to the stream function by: 
\begin{align}
\label{pepsi}
    \frac{\partial p}{\partial x} &= \frac{\partial}{\partial y} \nabla^2\psi, \\
     \frac{\partial p}{\partial y} &= -\frac{\partial}{\partial x} \nabla^2\psi. 
\end{align}

In the reference frame of the moving sheet, the dimensionless boundary conditions for the velocity are then: 
\begin{align}
\frac{\partial \psi}{\partial y}(x_{\textrm{s}},y_{\textrm{s}},t) &= \epsilon a\sin(x-t-\phi), \\
 \frac{\partial \psi}{\partial x}(x_{\textrm{s}},y_{\textrm{s}},t) &= \epsilon b\cos(x-t), \\
  \frac{\partial \psi}{\partial y}(x,d_0+\delta,t) &= U,\\
  \frac{\partial \psi}{\partial x}(x,d_0+\delta,t) &= -\frac{\textrm{d}\delta}{\textrm{d} t}(x,t),
\end{align}
where $U=k\bar{U}/\omega$ is the unknown dimensionless swimming velocity.

\section{\label{sec:perturbation_analysis}Perturbation analysis}

As has been done previously~\cite{taylor1951analysis,wu1961swimming,reynolds1965swimming,wu1971hydromechanics,katz1974propulsion}, we proceed with the calculation in the small-amplitude limit of the wave. To do so, we introduce a perturbative expansion in $\epsilon$. Since there is no flow in the absence of oscillation, the perturbative expansion starts at first order in $\epsilon$. Hence, we set: 
\begin{align}
    \psi &\simeq \epsilon \psi_1+\epsilon^2 \psi_2+O(\epsilon ^3), \\
    \Vec{u} &\simeq \epsilon \Vec{u}_{1}+\epsilon^2 \Vec{u}_{2}+O(\epsilon ^3), \\
    p &\simeq \epsilon p_{1}+\epsilon^2 p_{2}+O(\epsilon ^3), \\
    \delta &\simeq \epsilon \delta_1+\epsilon^2 \delta_{2}+O(\epsilon ^3),\\
     U &\simeq \epsilon U_1+\epsilon^2 U_{2}+O(\epsilon ^3).
    \end{align}

\subsection{First-order solution}
The governing equation for the dynamics of the fluid is the biharmonic equation: 
\begin{align}
    \nabla^4\psi_1 = 0, 
\end{align}
and it satisfies the following boundary conditions: 
\begin{align}
  \frac{\partial \psi_1}{\partial y}(x,0,t) &= a\sin(x-t-\phi), \label{eq:O1_bc1}\\
\frac{\partial \psi_1}{\partial x}(x,0,t) &= b\cos(x-t), \label{eq:O1_bc2}\\
   \frac{\partial \psi_1}{\partial y}(x,d_0,t)&= U_1, \label{eq:O1_bc3}\\
\frac{\partial \psi_1}{\partial x}(x,d_0,t)&= -\kappa\frac{\partial }{\partial t}\left[p_1(x,d_0,t)+2\frac{\partial^2 \psi_1}{\partial x \partial y}(x,d_0,t)\right]. \label{eq:O1_bc4}
\end{align}
The general $2\pi$-periodic solution of the biharmonic equation can be written down as: 
\begin{align}
    \psi_1 = K_1 y+ L_1 y^2+ M_1 y^3 + \sum_{n\geq1}\left\{(G_{1n}+yH_{1n})\cosh[n(y-d_0)]\right.\left.+(I_{1n}+yJ_{1n})\sinh[n(y-d_0)]\right\}\textrm{e}^{in(x-t)}, 
\end{align}
where the capital letters denote the complex coefficients to be determined by applying the boundary conditions. The coefficients $L_1$ and $M_1$ vanish since no mean shear or pressure gradient is applied in the system. Given the boundary conditions, we obtain that $K_1 = U_1 = 0$, and the series terminates at $n=1$. By referring to Eq.~(\ref{pepsi}), the pressure at the soft boundary turns out to be: 
\begin{align}
    p_1(x,d_0,t) = \mathcal{R}\left[-2iH_{11}\textrm{e}^{i(x-t)}\right], \label{eq:p1_wall}
\end{align}
where $\mathcal{R}$ denotes the real part. Thus, we end up with the following system of equations: 
\begin{align}
-(G_{11}+J_{11})\sinh d_0+(H_{11}+I_{11})\cosh d_0 &=-ia\textrm{e}^{-i\phi},\label{subeq:1}\\
G_{11}(\cosh d_0)+I_{11}(-\sinh d_0) &=-ib,\label{subeq:2}\\
H_{11}+I_{11}+d_0J_{11} &=0,\label{subeq:3}\\
G_{11}+ d_0 H_{11}-2i\kappa I_{11}-2i\kappa d_0 J_{11} &=0.\label{subeq:4}
\end{align}
The equations above can be solved to obtain the coefficients, as:
\begin{align}
	G_{11} &=-\frac{2 i \textrm{e}^{-i \phi} (d_0+2 i \kappa) \sinh (d_0) \left\{a d_0+b \textrm{e}^{i \phi} [d_0 \coth (d_0)+1]\right\}}{1+2d_0^2-\cosh (2d_0)+2i\kappa[2d_0+\sinh (2d_0)]},  \label{eq:G11}\\ 
    H_{11} &=\frac{2 i \left[b d_0 \cosh (d_0)+\sinh (d_0) \left(b+a d_0 \textrm{e}^{-i \phi}\right)\right]}{1+2d_0^2-\cosh (2d_0)+2i\kappa[2d_0+\sinh (2d_0)]},  \label{eq:H11}\\
    I_{11} &= -\frac{i \textrm{e}^{-i \phi} \left\{2 a d_0 (d_0+2 i \kappa) \cosh (d_0)+2 b \textrm{e}^{i \phi} \sinh (d_0) \left[d_0^2+2 i d_0 \kappa+d_0 \coth (d_0)+1\right]\right\}}{1+2d_0^2-\cosh (2d_0)+2i\kappa[2d_0+\sinh (2d_0)]},  \label{eq:I11}\\
    J_{11} &= \frac{2 i \textrm{e}^{-i \phi} \left\{\sinh (d_0) \left[-a+b \textrm{e}^{i \phi} (d_0+2 i \kappa)\right]+a (d_0+2 i \kappa) \cosh (d_0)\right\}}{1+2d_0^2-\cosh (2d_0)+2i\kappa[2d_0+\sinh (2d_0)]}. \label{eq:J11}
\end{align}

Finally, the deflection of the soft boundary is related to the normal stress by Eq.~(\ref{eq:winkler_nondim}). Hence, at first order in $\epsilon$, it can be computed using the above results, through:
\begin{align}
\label{wink1}
    \delta_1(x,t)  = \kappa \left[p_1(x,d_0,t)+2\frac{\partial^2 \psi_1}{\partial x \partial y}(x,d_0,t)\right]. 
\end{align}

\subsection{Second-order solution}
Since the swimming velocity at first order in $\epsilon$ is null ($U_1 = 0$), we need to analyse the dynamics at second order in $\epsilon$ to obtain the swimming velocity of the sheet. The governing equation for the dynamics of the fluid is once again:
\begin{align}
    \nabla^4\psi_2 = 0, 
\end{align}
with the following boundary conditions: 
\begin{align}
 \frac{\partial \psi_2}{\partial y}(x,0,t)&= -a\cos(x-t-\phi)\frac{\partial^2 \psi_1}{\partial x\partial y}(x,0,t) -b\sin(x-t)\frac{\partial^2 \psi_1}{\partial y^2}(x,0,t), \label{eq:O2_bc1}\\
\frac{\partial \psi_2}{\partial x}(x,0,t) &= -a\cos(x-t-\phi)\frac{\partial^2 \psi_1}{\partial x^2}(x,0,t) -b\sin(x-t)\frac{\partial^2 \psi_1}{\partial y\partial x}(x,0,t), \label{eq:O2_bc2}\\
\frac{\partial \psi_2}{\partial y}(x,d_0,t)&= U_2-\kappa \left[p_1(x,d_0,t)+2\frac{\partial^2 \psi_1}{\partial x \partial y}(x,d_0,t)\right]\frac{\partial^2\psi_1}{\partial y^2}(x,d_0,t), \label{eq:O2_bc3}\\
\frac{\partial \psi_2}{\partial x}(x,d_0,t) &= -\kappa\frac{\partial }{\partial t}\left[p_2(x,d_0,t)+2\frac{\partial^2 \psi_2}{\partial x \partial y}(x,d_0,t)\right]-\kappa \left[p_1(x,d_0,t)+2\frac{\partial^2 \psi_1}{\partial x \partial y}(x,d_0,t)\right]\frac{\partial^2\psi_1}{\partial y\partial x}(x,d_0,t). \label{eq:O2_bc4}
\end{align}
Note that we have removed the second-order convective part, $-\kappa U_1\partial_x\delta_1(x,d_0,t)$, of the time derivative in the last equation, since $U_1=0$. 
Once again, the stream function can be written down as: 
\begin{align}
    \psi_2 = K_2 y+ \{(G_{22}+yH_{22})\cosh[2(y-d_0)]+(I_{22}+yJ_{22})\sinh[2(y-d_0)]\}\textrm{e}^{2i(x-t)}, 
\end{align}
where only the $n=2$ terms are kept due to the right-hand sides of Eqs.~(\ref{eq:O2_bc1}-\ref{eq:O2_bc4}). The swimming speed of the sheet can be determined by averaging the right-hand side of Eq.~(\ref{eq:O2_bc1}) over one wavelength and equating it with the corresponding average of the right-hand side of Eq.~(\ref{eq:O2_bc3}). This leads to: 
\begin{align}
 U_2 &= \frac{f_1(a,b,\phi,d_0,\kappa)}{f_2(a,b,\phi,d_0,\kappa)}, \label{eq:U2}
\end{align}
where the auxiliary functions $f_1$ and $f_2$ are: 
\begin{align}
\begin{split}
 	 f_1 &= 4 \cosh (2 d_0) \left\{a^2 \left[d_0^2 \left(4 \kappa^2+2\right)+1\right]+a b \left(4 \kappa^2+1\right) \sinh (2 d_0) \cos (\phi)-2 a b d_0 \cos (\phi)+b^2 \left[4 \left(d_0^2-1\right) \kappa^2-1\right]\right\}\\
	 &-a^2 \left[8 d_0^2 \left(d_0^2+6 \kappa^2+1\right)-4 \kappa^2+3\right]-4 a \sinh (2 d_0) \left\{16 a d_0 \kappa^2+b \left[d_0^2 \left(2-8 \kappa^2\right)+4 \kappa^2+1\right] \cos (\phi)\right\} \\
	 &+32 a b \kappa \sin (\phi) \left[d_0^2 \cosh ^2 (d_0)-\sinh ^2 (d_0)\right]+8 a b d_0 \left(2 d_0^2+1\right) \cos (\phi)-\left(4 \kappa^2+1\right) (a-b) (a+b) \cosh (4 d_0)\\
	 &+b^2 \left[-8 d_0^4+4 \left(3-4 d_0^2\right) \kappa^2+3\right], 
\end{split}\\
 f_2 &= 2 \left[8 d_0^2 \left(d_0^2+4 \kappa^2+1\right)-4 \left(2 d_0^2+1\right) \cosh (2 d_0)+32 d_0 \kappa^2 \sinh (2 d_0)+\left(4 \kappa^2+1\right) \cosh (4 d_0)-4 \kappa^2+3\right]. 
\end{align}
\begin{figure}
     \begin{subfigure}[b]{0.45\textwidth}
         \caption{}
         \centering
         \includegraphics[width=8cm]{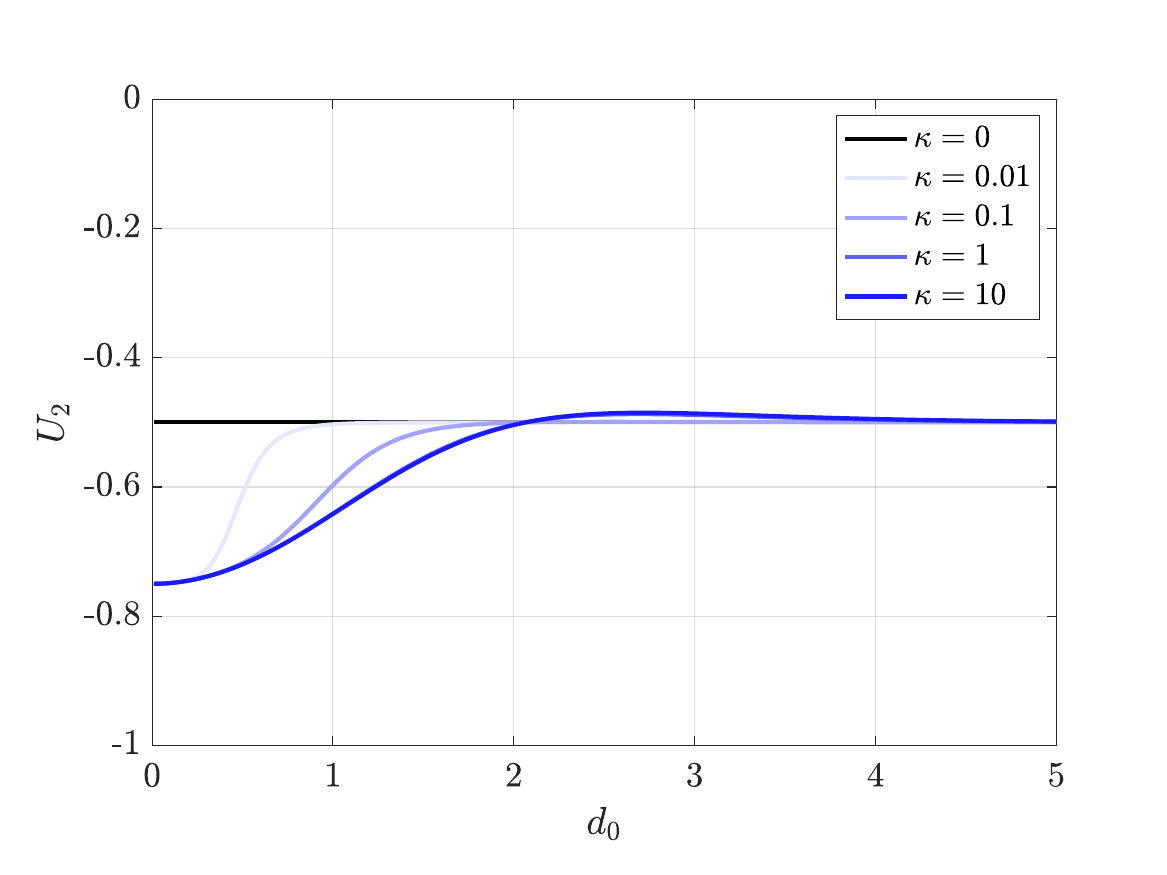}
          \label{fig:velocity_a_waves}
     \end{subfigure}
     \hfill
     \begin{subfigure}[b]{0.45\textwidth}
        \caption{}
         \centering
         \includegraphics[width=8cm]{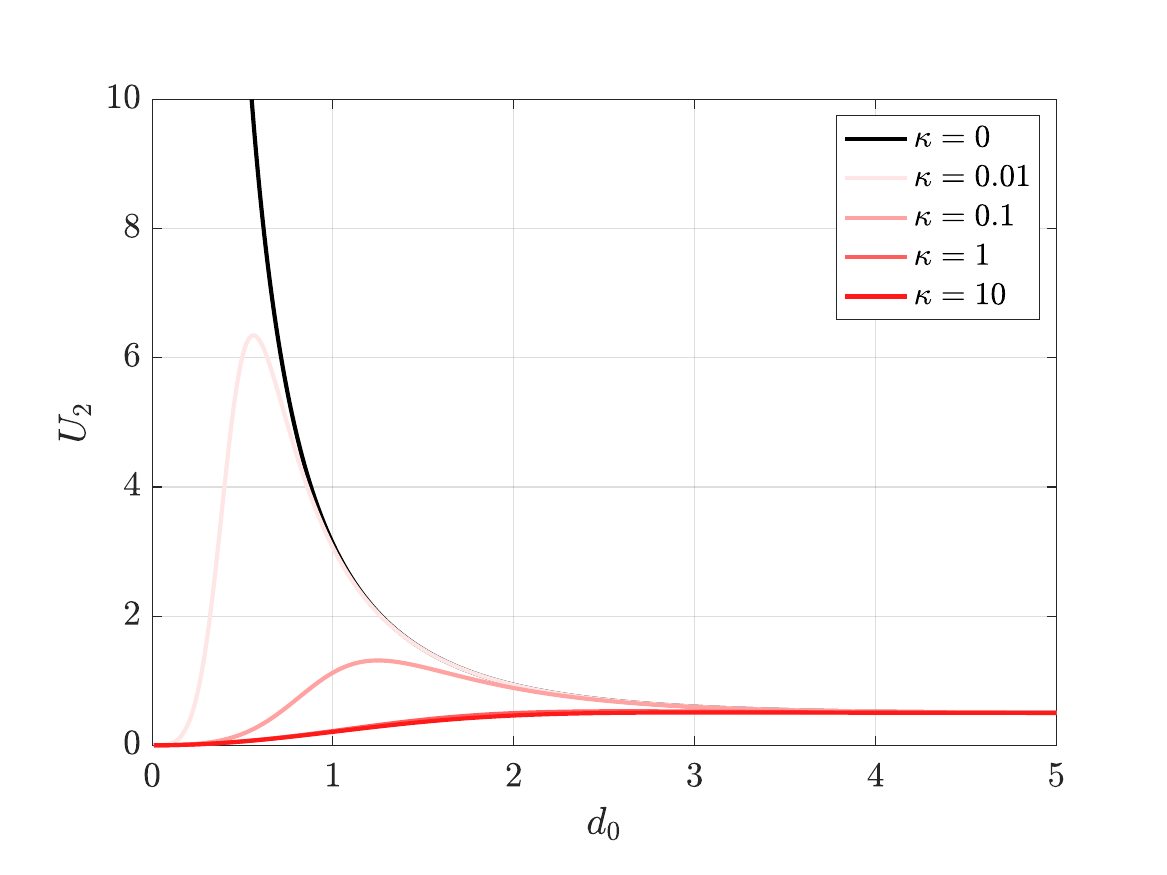}
          \label{fig:velocity_b_waves}
     \end{subfigure}
    \caption{Dimensionless second-order swimming velocity $U_2$ of the  sheet as a function of the dimensionless gap distance $d_0$ between the sheet and the soft boundary, for several values of the dimensionless compliance $\kappa$ as indicated, calculated by using Eq.~(\ref{eq:U2}) for: (a) longitudinal waves only ($a=1$, $\phi=0$, $b = 0$); (b) transverse waves only ($a = 0, b=1$). The black curves correspond to the case of a rigid boundary~\cite{reynolds1965swimming,katz1974propulsion}, for comparison.}
        \label{fig:U2}
\end{figure}

To gain better intuition about the dependencies of the swimming velocity on the various parameters, we consider individual waving motions. The velocity variations are plotted in Fig.~\ref{fig:U2} as functions of the gap distance $d_0$ for different values of the dimensionless compliance $\kappa$. The black lines correspond to the case of an undeformable boundary, as a reference. For longitudinal waves, the swimming velocity is mainly negative, and increases further in magnitude as the softness of the wall increases (see Fig.~\ref{fig:U2}a). Besides, as the distance between the sheet and the soft boundary reduces, the velocity increases in magnitude, and this effect is more potent at larger gaps for larger values of the compliance. However, the magnitudes of the velocity for all $\kappa$ values all saturate to the same value as the gap vanishes, with a consistent increase by $\sim50\%$. On the other hand, for transverse waves, the velocity is positive. When the wall is rigid, the velocity magnitude diverges as the gap is reduced. In this case, the velocity for small $d_0$ is inversely proportional to $d_0^{\,2}$. However, this divergence is tamed by the introduction of even a small amount of wall elasticity, as seen in Fig.~\ref{fig:U2}b where the velocity values of the swimming sheet all converge to $0$ at vanishing gap when $\kappa \neq 0$. In this case, at vanishing $d_0$, the velocity now increases as $\sim d_0^2$. 

For both the cases of longitudinal and transverse waves, there are limiting behaviours for both $\kappa \rightarrow 0$ and $\kappa \rightarrow \infty$, which can be obtained by studying the asymptotics of Eq.~(\ref{eq:U2}). In the $\kappa \rightarrow \infty$ limit for longitudinal waves, one has: 
\begin{align}
\left.U_2\right|_{\kappa \rightarrow \infty} = -\frac{a^2 \left[12 d_0^2-4 d_0^2 \cosh (2 d_0)+16 d_0 \sinh (2 d_0)+\cosh (4 d_0)-1\right]}{4 [2 d_0+\sinh (2 d_0)]^2},
\end{align} 
and, in the $\kappa \rightarrow \infty$ limit for transverse waves, one has: 
\begin{align}
\left.U_2\right|_{\kappa \rightarrow \infty} = \frac{b^2 \sinh ^2 (d_0) \left[2 d_0^2+\cosh (2 d_0)-1\right]}{[2 d_0+\sinh (2 d_0)]^2}.
\end{align}
These functions match the large-$\kappa$ solutions (not shown). It is interesting to note that the swimming velocity for a highly-deformable boundary becomes essentially independant of the elastic compliance of the wall. In the other limiting behavior, \textit{i.e.} $\kappa \rightarrow 0$, we obtain the same results (see Fig.~\ref{fig:U2}) as in previous studies for rigid boundaries~\cite{reynolds1965swimming,katz1974propulsion}. 

\begin{figure}[h]
\begin{center}
\includegraphics[width=8cm]{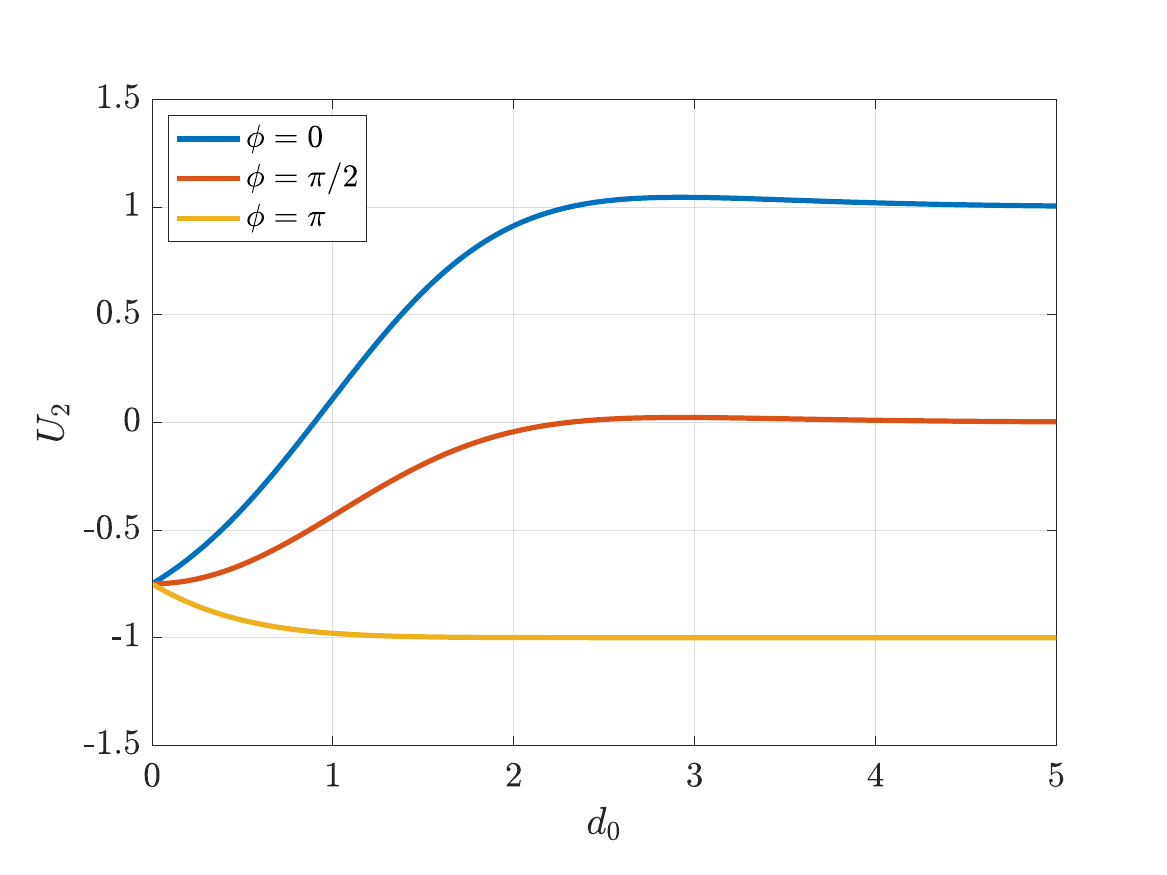}
\caption{Dimensionless second-order swimming velocity $U_2$ of the sheet as a function of the dimensionless gap distance $d_0$ between the sheet and the soft boundary, for three values of the phase shift $\phi$ between the longitudinal and transverse waves, as calculated from Eq.~(\ref{eq:U2}), for $\kappa \rightarrow \infty$, $a = 1$ and $b=1$. }
\label{fig:U2_ab_waves_kappainf}
\end{center}
\end{figure}

When both $a$ and $b$ are not zero, the velocity also depends on the phase difference $\phi$ between the transverse and longitudinal waves. For illustration purposes, we focus on the situation where the boundary is extremely soft ($\kappa \rightarrow \infty$). We plot in Fig.~\ref{fig:U2_ab_waves_kappainf} the swimming velocity in this limit, as a function of the gap distance $d_0$, for $a = b = 1$, and for different values of $\phi$. We see that for large gaps, increasing the phase difference $\phi$ reduces the velocity and can change its sign, in agreement with Taylor's result~\cite{taylor1951analysis}. In contrast, for small gaps, the velocity is always negative, indicating the importance of longitudinal waves near a soft boundary. 

\subsection{Leading-order energetics}
We now focus on the energetics of the system art leading-order in $\epsilon$, specifically exploring the rate at which work is done on the fluid. Since the power varies quadratically in velocity, the perturbation expansion in $\epsilon$ for the power (per unit length) $\dot {W}$ can be written as: 
\begin{align}
\dot{W} \simeq \epsilon^2\dot{W}_2+O(\epsilon^4).
\end{align}
Over one wavelength, the leading-order power (per unit length) reads: 
\begin{align}
\dot{W}_2 = &-\int_0^{2\pi} u_{1,x}\sigma_{1,xy}|_{y = 0}\textrm{d}x-\int_0^{2\pi} u_{1,y}\sigma_{1,yy}|_{y = 0}\textrm{d}x+\int_0^{2\pi} u_{1,x}\sigma_{1,xy}|_{y = d_0}\textrm{d}x+\int_0^{2\pi} u_{1,y}\sigma_{1,yy}|_{y = d_0}\textrm{d}x,
\end{align}
where the components of the stress tensor $\sigma$ are given by: 
\begin{align}
\sigma_{1,xy} &= \frac{\partial u_{1,x}}{\partial y}+ \frac{\partial u_{1,y}}{\partial x},\\
\sigma_{1,yy} &= -p_1+ 2\frac{\partial u_{1,y}}{\partial y}. 
\end{align}
Using the coefficients calculated previously for the stream function $\psi_{1}$, the above expressions can be used to calculate the leading-order power, as:
\begin{align}
\dot{W}_2 &= \frac{g_1(a,b,\phi,d_0,\kappa)}{g_2(a,b,\phi,d_0,\kappa)}, \label{eq:W2}
\end{align}
where:
\begin{align}
\begin{split}
g_1(a,b,\phi,d_0,\kappa) &= 4 \pi  \left[-2 d_0^2+\cosh (2 d_0)-1\right] \left\{2 d_0 \left[-a^2+2 a b d_0 \cos (\phi)+b^2\right]+\left(a^2+b^2\right) \sinh (2 d_0)\right\}\\
&+16 \pi  \kappa^2 [2 d_0+\sinh (2 d_0)]\left[\left(a^2+b^2\right) \cosh (2 d_0)+a^2-4 a b d_0 \cos (\phi)-b^2\right],
\end{split}\\
g_2(a,b,\phi,d_0,\kappa) &= 2 \left[-2 d_0^2+\cosh (2 d_0)-1\right]^2+8 \kappa^2 \left[2 d_0+\sinh (2 d_0)\right]^2. 
\end{align}
\begin{figure}
     \begin{subfigure}[b]{0.45\textwidth}
         \caption{}
         \centering
         \includegraphics[width=8cm]{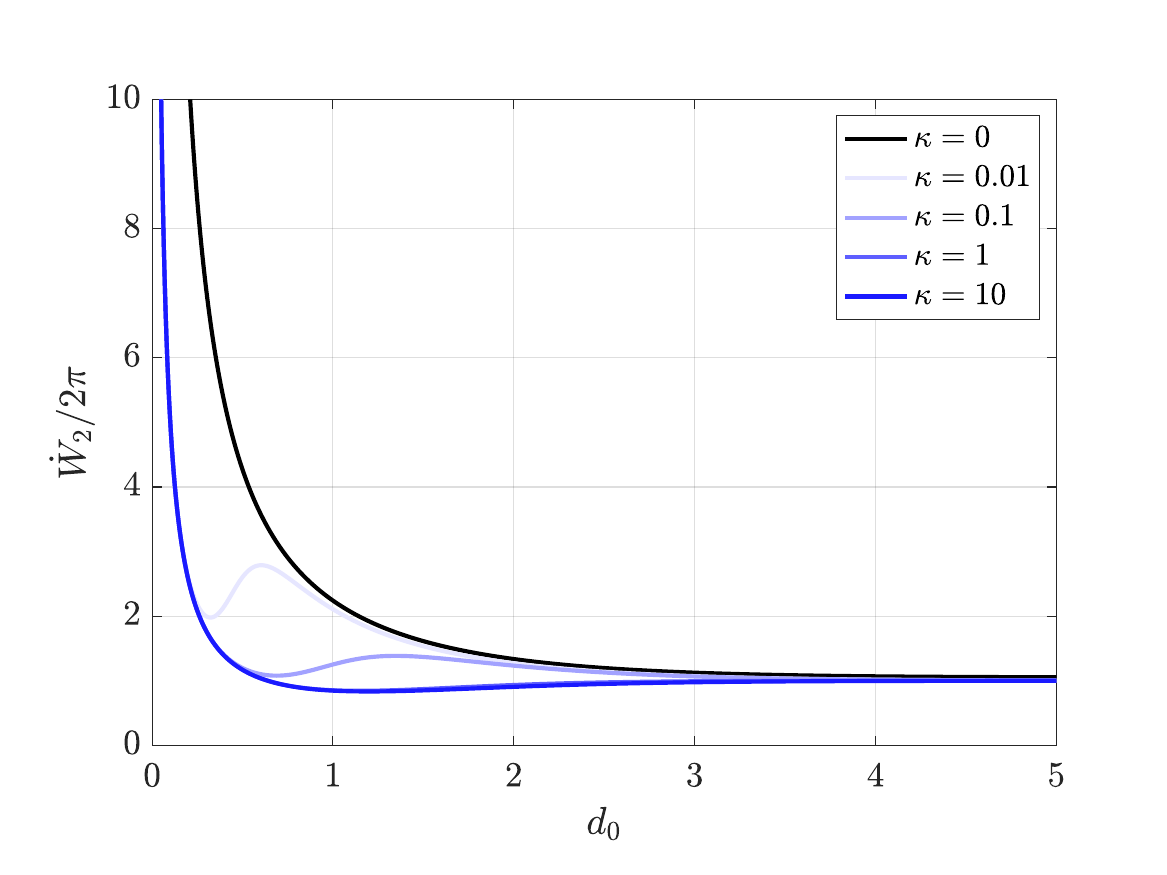}
          \label{fig:rate_of_work_a_waves}
     \end{subfigure}
     \hfill
     \begin{subfigure}[b]{0.45\textwidth}
        \caption{}
         \centering
         \includegraphics[width=8cm]{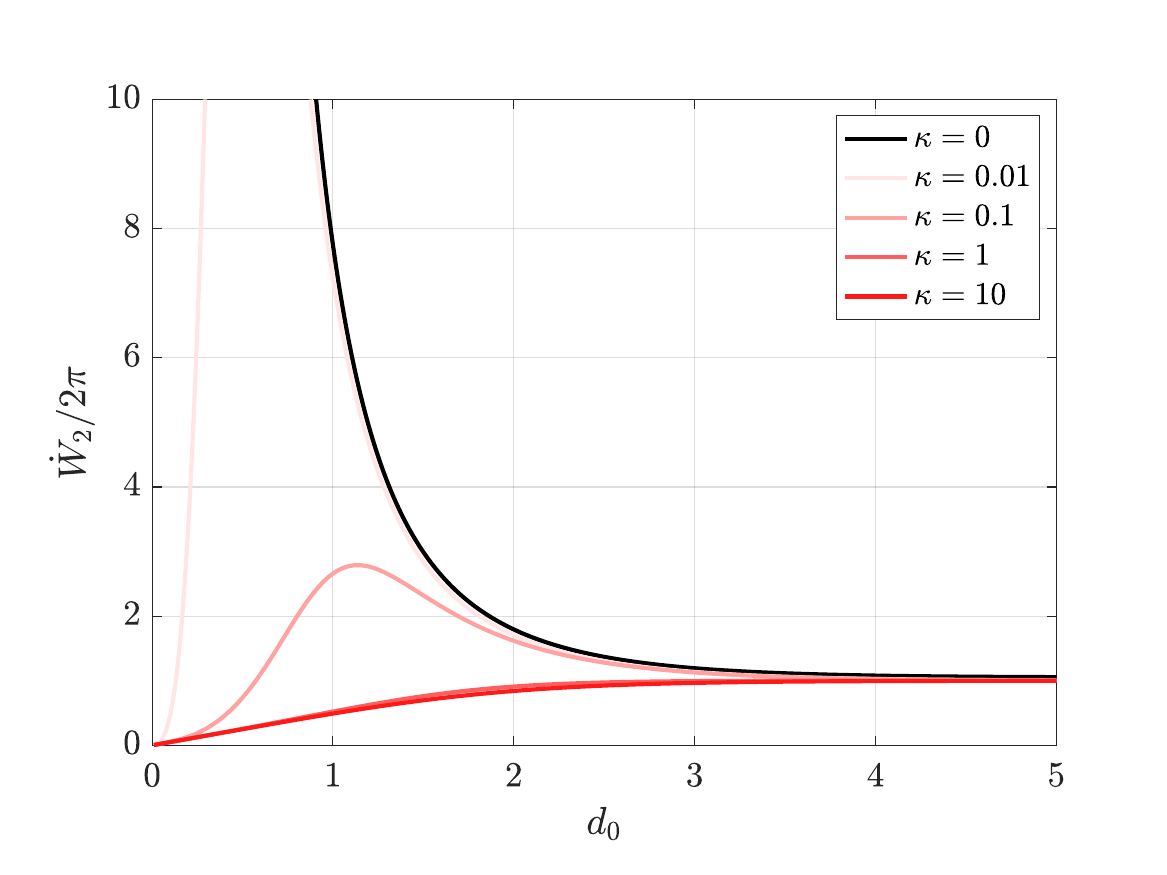}
          \label{fig:rate_of_work_b_waves}
     \end{subfigure}
    \caption{Dimensionless leading-order power $\dot{W}_2$ (normalized by $2\pi$) as a function of the dimensionless gap distance $d_0$ between the sheet and the soft boundary, for several values of the dimensionless compliance $\kappa$, as calculated from Eq.~(\ref{eq:W2}) for: (a) longitudinal waves only ($a=1$, $\phi=0$, $b = 0$); (b) transverse waves only ($a = 0$, $b=1$). The black curves correspond to the case of a rigid boundary. }
        \label{fig:W2}
\end{figure}

The results for the power of individual waving motions are shown in Fig.~\ref{fig:W2}. The leading-order power near a rigid boundary diverges for both longitudinal and transverse waves as the gap vanishes ($d_0 \rightarrow 0$). For longitudinal waves, the behaviour of the leading-order power at vanishing gaps remains inversely proportional to $d_0$, with the power being always smaller near a softer wall. For transverse waves, the divergence of the power is tamed as soon as elasticity of the boundary is introduced. Furthermore, in the latter case, similarly to the swimming velocity, the power vanishes as the gap between the sheet and the soft boundary tends to zero. For the extreme cases of $\kappa \rightarrow 0$ and $\kappa \rightarrow \infty$, the curves indicate clear limiting behaviours. For $\kappa \rightarrow \infty$, the limiting behavior can be obtained by asymptotic expansion of Eq.~(\ref{eq:W2}), which for longitudinal waves reads: 
\begin{align}
\left.\dot{W}_2\right|_{\kappa \rightarrow \infty} = \frac{2 \pi  a^2 \cosh ^2 (d_0)}{d_0+\sinh (d_0) \cosh (d_0)},
\end{align}
while for transverse waves it reads: 
\begin{align}
\left.\dot{W}_2\right|_{\kappa \rightarrow \infty} = \frac{2 \pi  b^2 \sinh ^2 (d_0)}{d_0+\sinh (d_0) \cosh (d_0)}.
\end{align}
For a composite wave, it becomes:
\begin{align}
 \left.\dot{W}_2\right|_{\kappa \rightarrow \infty} = \frac{2 \pi  \left[\left(a^2+b^2\right) \cosh (2 d_0)+a^2-4 a b d_0 \cos (\phi)-b^2\right]}{2 d_0+\sinh (2 d_0)}.
\end{align}

\begin{figure}[h]
\begin{center}
\includegraphics[width=8cm]{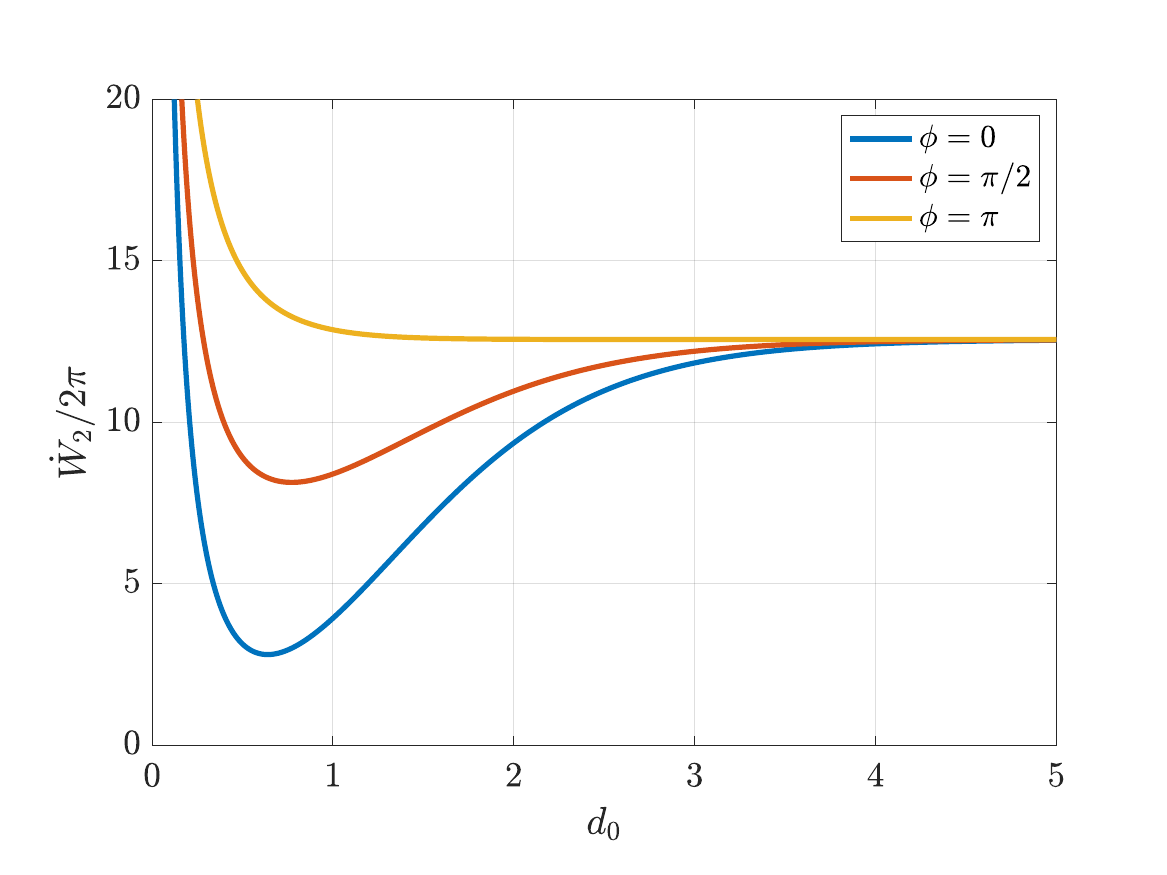}
\caption{Dimensionless leading-order power $\dot{W}_2$ (normalized by $2\pi$) as a function of the dimensionless gap distance $d_0$ between the sheet and the soft boundary, for several values of the phase shift $\phi$ between the longitudinal and transverse waves, as calculated from Eq.~(\ref{eq:W2}), for $\kappa \rightarrow \infty$, $a =1$ and $b=1$.  }
\label{fig:W2_ab_waves_kappinf}
\end{center}
\end{figure}

When both $a$ and $b$ are not zero, the power decreases for increasing softness but still diverges at vanishing gaps, owing to the divergence observed in the case of longitudinal waves only. Furthermore, the power now depends on the phase difference $\phi$ between the two waving modes. For extremely soft boundaries, where $\kappa \rightarrow \infty$, we plot in Fig.~\ref{fig:W2_ab_waves_kappinf} the power as a function of the gap distance, for several phase differences. The result indicates that increasing the phase difference leads to an increasing power. 

\subsection{Deformation of the soft boundary}
At order one in $\epsilon$, we assume that the wall deforms along $y$ with an amplitude $A_{\textrm{w}}$ and a phase delay $\theta$ with respect to the applied transverse wave, \textit{i.e.}: 
\begin{align}
\delta_1 = A_{\textrm{w}} \sin( x-t+\theta). \label{eq:delta_1}
\end{align}
By invoking Eq.~(\ref{wink1}), one thus gets: 
\begin{align}
A_{\textrm{w}} & = 4\kappa\left[\frac{h_1(a,b,\phi,d_0,\kappa)}{h_2(a,b,\phi,d_0,\kappa)}\right]^{1/2},\\
\theta & = \tan^{-1}\left[\frac{h_3(a,b,\phi,d_0,\kappa)}{h_4(a,b,\phi,d_0,\kappa)}\right], 
\end{align}
where: 
\begin{align}
h_1 & = 2 a^2 d_0^2 \sinh ^2 (d_0) +4 a b d_0 \sinh (d_0) \cos (\phi) [\sinh (d_0)+d_0 \cosh (d_0)]+2 b^2 [\sinh (d_0)+d_0 \cosh (d_0)]^2,\\
h_2 & = 2 \left[-2 d_0^2+\cosh (2 d_0)-1\right]^2+8 \kappa^2 [2 d_0+\sinh (2 d_0)]^2,\\
h_3 & = \left[2 d_0^2-\cosh (2 d_0)+1\right] \{\sinh (d_0) [a d_0 \cos (\phi)+b]+b d_0 \cosh (d_0)\}-4 a d_0 \kappa \sinh (d_0) \sin (\phi) [d_0+\sinh (d_0) \cosh (d_0)], \\
h_4 & = 4 \kappa [d_0+\sinh (d_0) \cosh (d_0)] \{\sinh (d_0) [a d_0 \cos (\phi)+b]+b d_0 \cosh (d_0)\}+a d_0 \sinh (d_0) \left[2 d_0^2-\cosh (2 d_0)+1\right] \sin (\phi).
\end{align}

\begin{figure}
     \begin{subfigure}[b]{0.45\textwidth}
         \caption{}
         \centering
         \includegraphics[width=8cm]{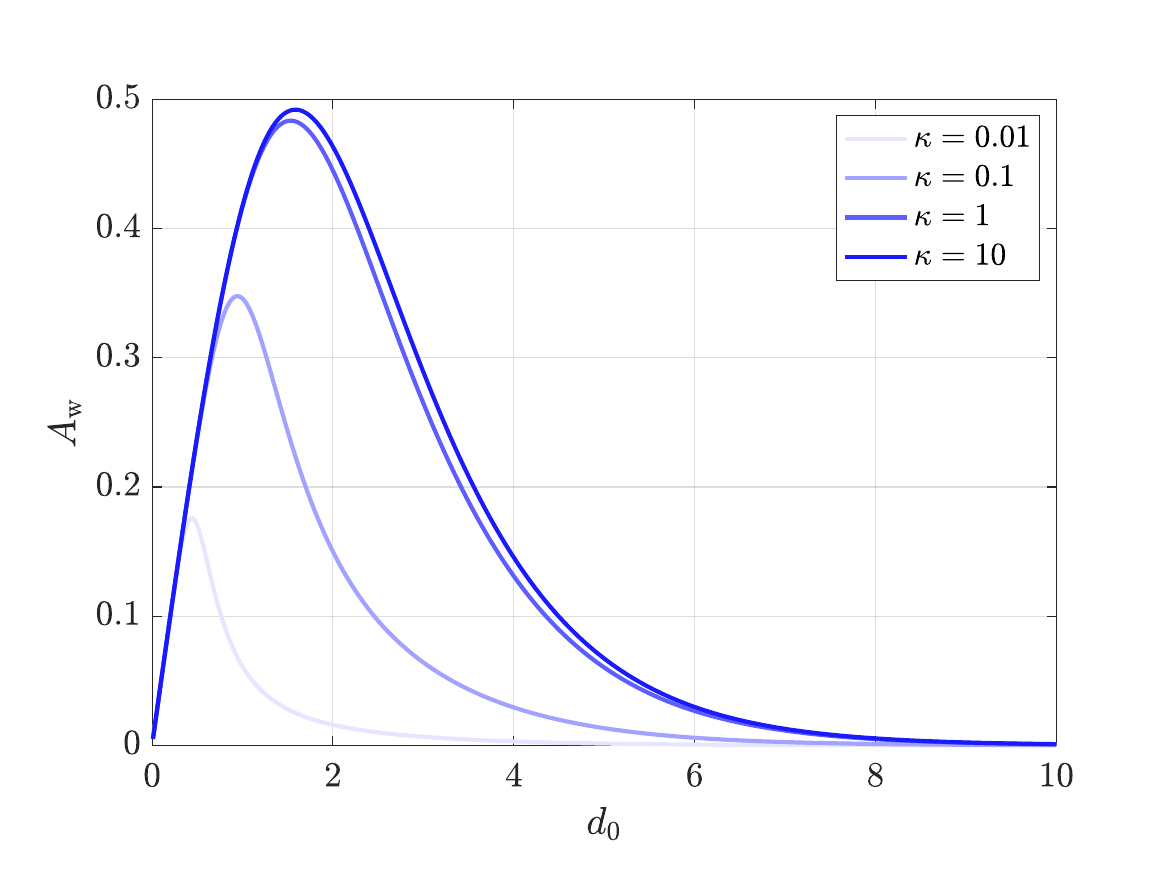}
          \label{fig:deformation_amplitude_a_waves}
     \end{subfigure}
     \hfill
     \begin{subfigure}[b]{0.45\textwidth}
        \caption{}
         \centering
         \includegraphics[width=8cm]{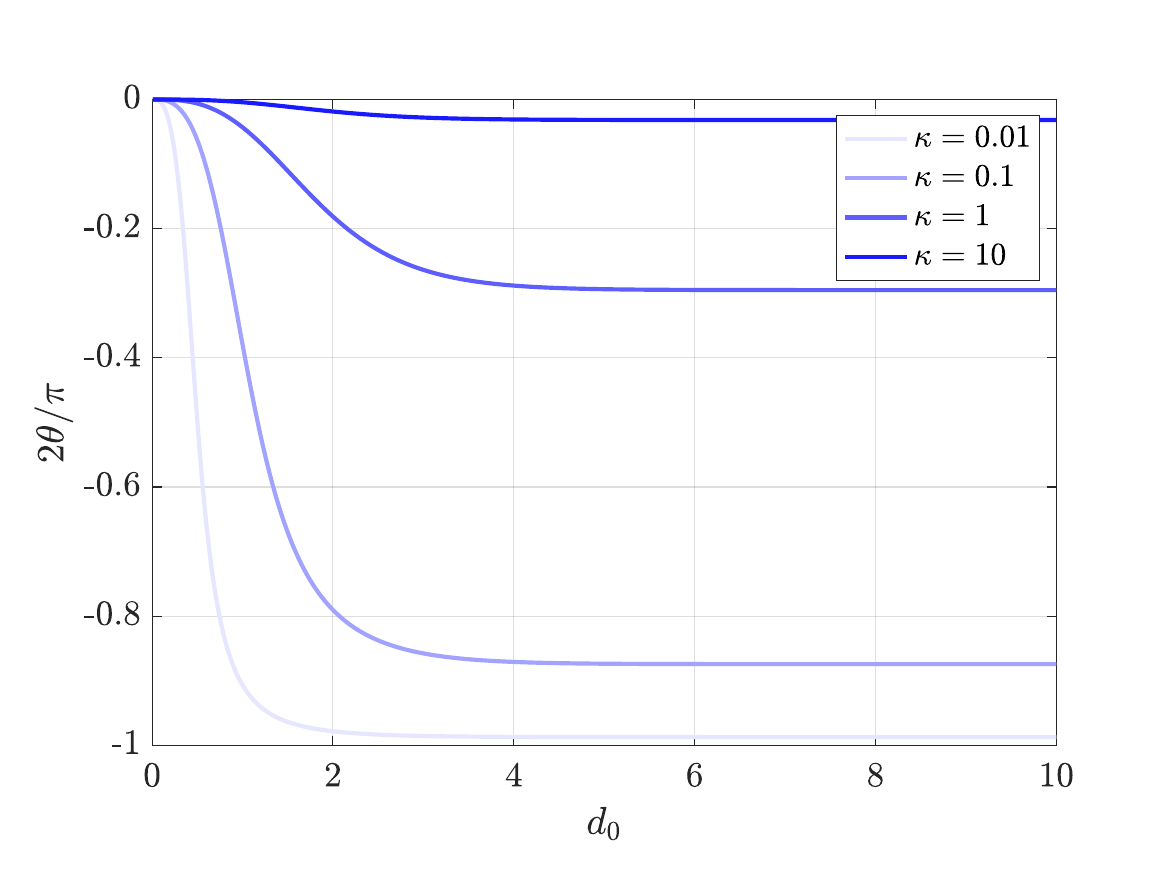}
          \label{fig:phase_offset_a_waves}
     \end{subfigure}
    \caption{(a) Dimensionless amplitude $A_{\textrm{w}}$ of the first-order deformation of the soft substrate as a function of the dimensionless gap distance $d_0$, for various values of the dimensionless compliance $\kappa$, as obtained from Eqs.~(\ref{wink1}) and~(\ref{eq:delta_1}) for a longitudinal wave with $a = 1$, $\phi=0$ and $b = 0$. (b) Phase delay $\theta$ (normalized by $\pi/2$) of the first-order deformation of the soft substrate as a function of the dimensionless gap distance $d_0$, in the same conditions as in (a).}
        \label{fig:deformation_a_waves}
\end{figure}
\begin{figure}
     \begin{subfigure}[b]{0.45\textwidth}
         \caption{}
         \centering
         \includegraphics[width=8cm]{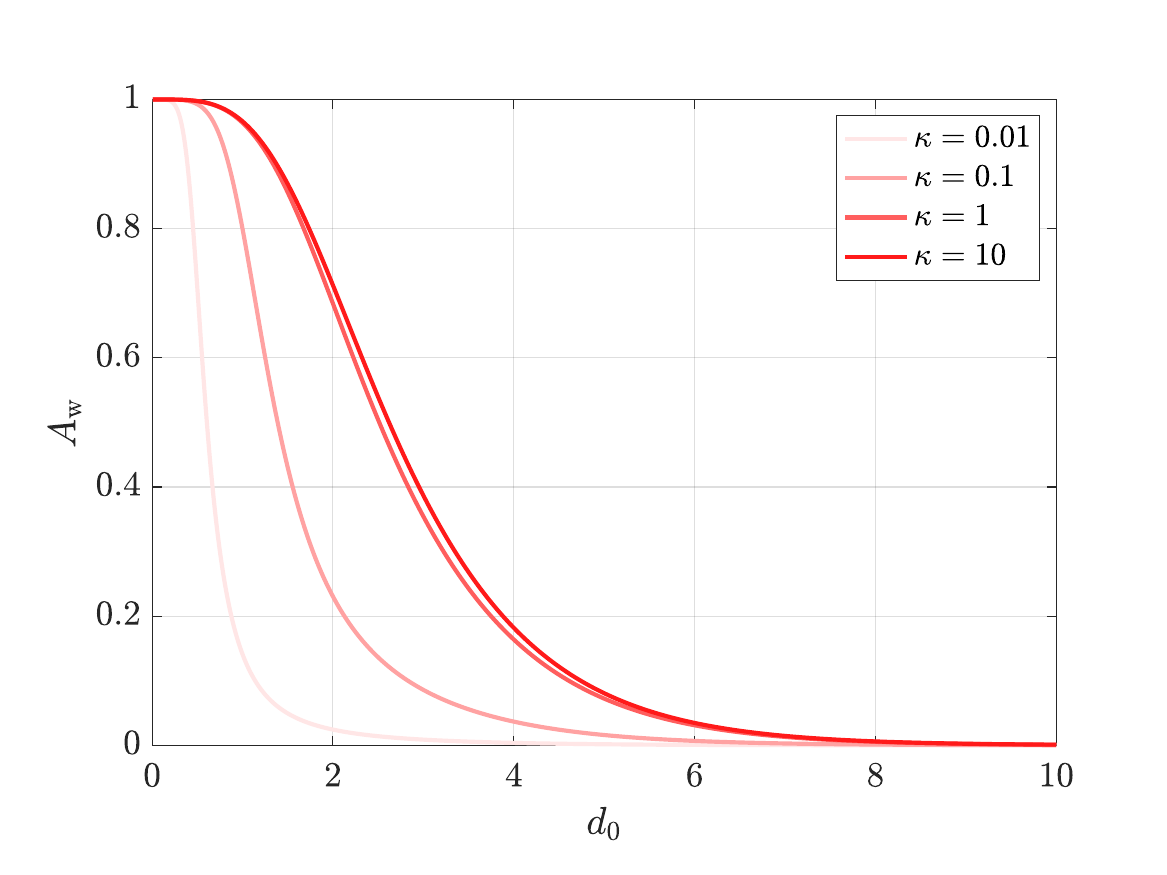}
          \label{fig:deformation_amplitude_b_waves}
     \end{subfigure}
     \hfill
     \begin{subfigure}[b]{0.45\textwidth}
        \caption{}
         \centering
         \includegraphics[width=8cm]{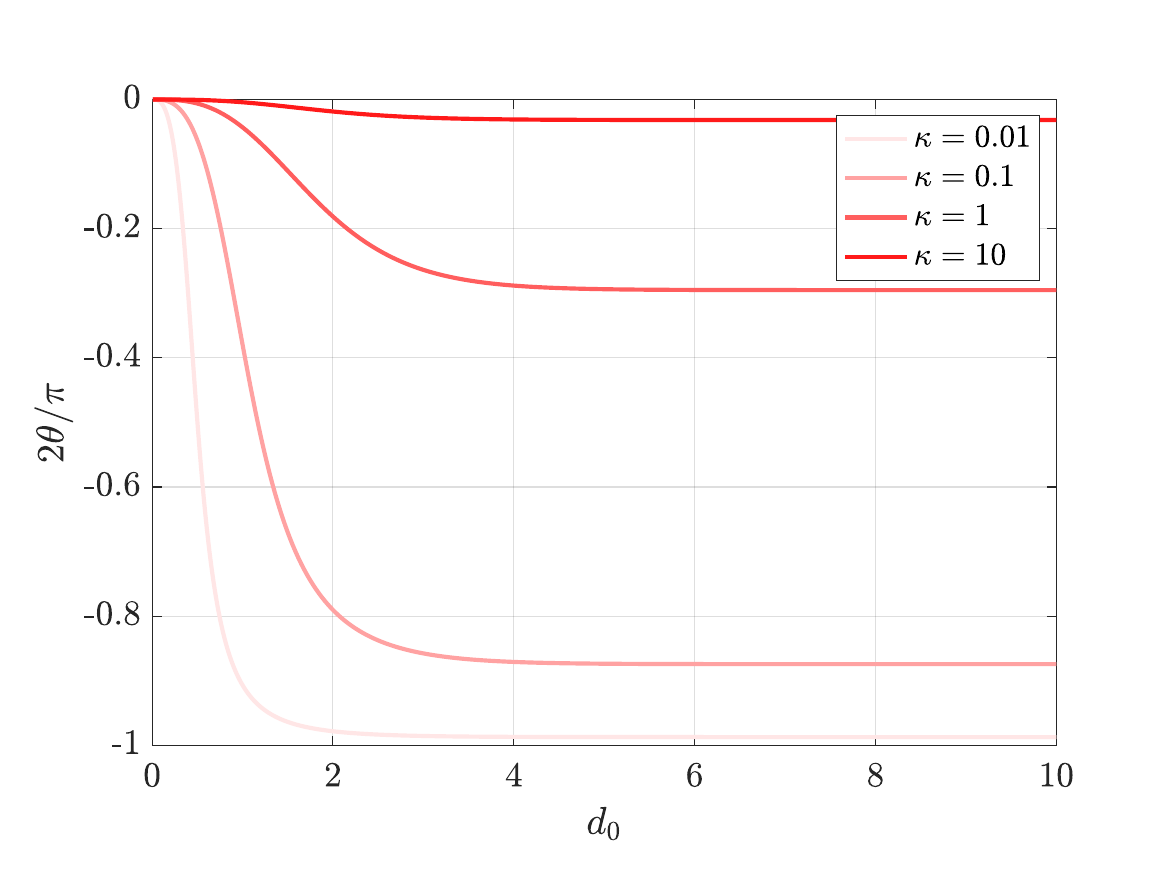}
          \label{fig:phase_offset_b_waves}
     \end{subfigure}
    \caption{(a) Dimensionless amplitude $A_{\textrm{w}}$ of the first-order deformation of the soft substrate as a function of the dimensionless gap distance $d_0$, for various values of the dimensionless compliance $\kappa$, as obtained from Eqs.~(\ref{wink1}) and~(\ref{eq:delta_1}) for a transverse wave with $a = 0$ and $b = 1$. (b) Phase delay $\theta$ (normalized by $\pi/2$) of the first-order deformation of the soft substrate as a function of the dimensionless gap distance $d_0$, in the same conditions as in (a).}
        \label{fig:deformation_b_waves}
\end{figure}

The deformation amplitudes and phase delays are plotted using the above equations for pure longitudinal waves in Fig.~\ref{fig:deformation_a_waves}, and for pure transverse waves in Fig.~\ref{fig:deformation_b_waves}, as functions of the gap distance and for several values of the compliance. The deformation amplitudes are starkly different for longitudinal waves and transverse waves passing through the sheet. In the case of longitudinal waves, as the gap distance between the sheet and the substrate reduces, the deformation amplitude reaches a maximum and entirely vanishes with vanishing gap distance. In contrast, it keeps on increasing monotonically with decreasing gap distance in the case of transverse waves. Moreover, the maximum deformation amplitude achieved in the case of longitudinal waves peaks at about half the amplitude of the applied wave. In contrast, the maximum deformation amplitude has the same amplitude as the applied wave in the case of transverse waves. 
Despite the dissimilar behaviors regarding the amplitude, the phase delays show similar behaviors. Indeed, in both cases, the magnitude of the phase delay reduces with increasing compliance, and further vanishes as the gap distance vanishes. 

These results reveal a complete softness-induced synchronization in the case of transverse waves, where the wall deforms with the same amplitude as the applied wave and no phase delay.  The results may also explain the variation of the leading-order power, which vanishes for transverse waves but diverges to infinity for longitudinal waves at vanishing gaps $d_0$. 

Finally, for the amplitude of deformation $A_{\textrm{w}}$, we see saturating behaviors at large values of $\kappa$, which for longitudinal waves only can be written down as: 
\begin{align}
\left.A_{\textrm{w}}\right|_{\kappa \rightarrow \infty} = \frac{2 ad_0\sinh (d_0)}{2 d_0+\sinh (2 d_0)},
\end{align}
and for transverse waves only, as: 
\begin{align}
\left.A_{\textrm{w}}\right|_{\kappa \rightarrow \infty}  = \frac{2 b[d_0 \cosh (d_0)+\sinh (d_0)]}{2 d_0+\sinh (2 d_0)}.
\end{align} 

\section{Conclusion}
In summary, we explored the influence of a nearby elastic boundary onto a model swimmer based on Taylor's idea of a sheet with transverse and longitudinal waves passing through it. For small-amplitude waves, the swimming velocity is tuned by the boundary's elasticity and the swimmer's distance from the wall, with contrasting characteristics depending on the nature of the waves. While transverse waves show a reduced swimming speed near soft boundaries, longitudinal waves lead to a larger swimming speed than that possible near a perfectly rigid boundary. Moreover, the introduction of elasticity controls the divergent nature of the swimming speed for transverse waves. We also calculated the exerted power and the deformation of the elastic boundary at leading orders in wave amplitude, which showed that a larger softness leads to an increased synchronization between the sheet and the elastic boundary. Furthermore, while the deformation wave induced in the soft boundary synchronized with the applied wave entirely in the case of transverse waves, the deformation caused by longitudinal waves was ultimately arrested as the gap distance was reduced towards zero. 
Future work could extend the current model by focusing on: i) a semi-infinite viscoelastic elastic solid, and ii) a finite-size swimmer, in order to delve deeper into the understanding and optimization of the gliding motion exhibited by certain microorganisms. 

\section{acknowledgments}
The authors acknowledge financial support from the European Union through the European Research Council under EMetBrown (ERC-CoG-101039103) grant. Views and opinions expressed are however those of the authors only and do not necessarily reflect those of the European Union or the European Research Council. Neither the European Union nor the granting authority can be held responsible for them. The authors also acknowledge financial support from the Agence Nationale de la Recherche under Softer (ANR-21-CE06-0029), and Fricolas (ANR-21-CE06-0039) grants. Finally, they thank the Soft Matter Collaborative Research Unit, Frontier Research Center for Advanced Material and Life Science, Faculty of Advanced Life Science at Hokkaido University, Sapporo, Japan.  

\bibliography{Jha2024}
\newpage
\end{document}